\documentclass[9pt,twocolumn,twoside]{pnas-new}
\usepackage{xr-hyper}
\usepackage{hyperref}
\templatetype{pnasresearcharticle} 
\title{Minimal vertex model explains how the amnioserosa avoids fluidization during \textit{Drosophila} dorsal closure}

\author[a,b,1]{Indrajit Tah}
\author[c,d,1]{Daniel Haertter} 
\author[e]{Janice M. Crawford}
\author[e,2]{Daniel P. Kiehart}
\author[d,2]{Christoph F. Schmidt}
\author[b,2]{Andrea J. Liu}

\affil[a]{Speciality Glass Division, CSIR-Central Glass and Ceramic Research Institute, Kolkata, India}
\affil[b]{Department of Physics and Astronomy, University of Pennsylvania, PA, USA}
\affil[c]{Institute of Pharmacology and Toxicology, University Medical Center and Campus Institute Data Science (CIDAS), University of Göttingen, Germany}
\affil[d]{Department of Physics and Soft Matter Center, Duke University, Durham, NC, USA}
\affil[e]{Department of Biology, Duke University, Durham, NC, USA}

\leadauthor{Tah, Haertter}
\significancestatement{During embryogenesis, cells in tissues can undergo significant shape changes. Many epithelial tissues fluidize, i.e. cells exchange neighbors, when the average cell aspect ratio increases above a threshold value, consistent with the standard vertex model. During dorsal closure in \textit{Drosophila melanogaster}, however, the amnioserosa tissue remains solid even as the average cell aspect ratio increases well above threshold. We introduce perimeter polydispersity and allow the preferred cell perimeters, usually held fixed in vertex models, to decrease linearly with time as seen experimentally. With these extensions to the standard vertex model, we capture experimental observations quantitatively. Our results demonstrate that vertex models can describe the behavior of the amnioserosa in dorsal closure by allowing normally fixed parameters to vary with time.}
\authorcontributions{All authors conceived and designed the research project. IT and DH performed the vertex model simulations and analyzed and visualized the data from model and experiments. JC and DPK contributed experimental data. All authors contributed to data interpretation and collaborated on writing the manuscript.}
\authordeclaration{All authors declare they have no competing interests.}
\equalauthors{\textsuperscript{1} Authors contributed equally. \textsuperscript{2} Authors contributed equally.}
\correspondingauthor{\textsuperscript{2} To whom correspondence should be addressed. E-mail: dkiehart@duke.edu, cfschmidt@phy.duke.edu, ajliu@physics.upenn.edu}

\begin{abstract} 
Dorsal closure is a process that occurs during embryogenesis of \textit{Drosophila melanogaster}. During dorsal closure, the amnioserosa (AS), a one-cell thick epithelial tissue that fills the dorsal opening, shrinks as the lateral epidermis sheets converge and eventually merge. During this process, the aspect ratio of amnioserosa cells increases markedly. The standard 2-dimensional vertex model, which successfully describes tissue sheet mechanics in multiple contexts, would in this case predict that the tissue should fluidize via cell neighbor changes. Surprisingly, however, the amnioserosa remains an elastic solid with no such events. We here present a minimal extension to the vertex model that explains how the amnioserosa can achieve this unexpected behavior. We show that continuous shrinkage of the preferred cell perimeter and cell perimeter polydispersity lead to the retention of the solid state of the amnioserosa. Our model accurately captures measured cell shape and orientation changes and predicts non-monotonic junction tension that we confirm with laser ablation experiments.
\end{abstract}

\dates{This manuscript was compiled on \today}
\doi{\url{www.pnas.org/cgi/doi/10.1073/pnas.XXXXXXXXXX}}

\begin{document}

\maketitle
\thispagestyle{firststyle}
\ifthenelse{\boolean{shortarticle}}{\ifthenelse{\boolean{singlecolumn}}{\abscontentformatted}{\abscontent}}{}
\dropcap{T}he developmental stage of dorsal closure in \textit{Drosophila melangaster} occurs roughly midway through embryogenesis and provides a model for cell sheet morphogenesis \cite{harden_signaling_2002,hayes_drosophila_2017,doi:10.1146/annurev-cellbio-111315-125357,aristotelous_mathematical_2018}. The amnioserosa (AS) consists of a single sheet of cells that fills a gap on the dorsal side of the embryo separating two lateral epidermal cell sheets. During closure, the AS shrinks in total area, driven by non-muscle myosin II acting on arrays of actin filaments in both the AS and actomyosin-rich cables in the leading edge of the lateral epidermis \cite{young_morphogenesis_1993,kiehart_multiple_2000,doi:10.1126/science.1079552,franke_nonmuscle_2005}. Ultimately, the AS disappears altogether. The entire closure process is choreographed by a developmental program that mediates changes in AS cell shapes as well as forces on adherens junctions between cells~\cite{pmid30399339,pmid31251912}. 

One might naively expect cells in the AS, which are glued to their neighbors by molecules such as E-cadherin, to maintain their neighbors, so that the tissue behaves like a soft, elastic solid even as it is strongly deformed by the forces driving dorsal closure. However, the time scale for making and breaking molecular bonds between cells (ms) is far faster than the time scale for dorsal closure (hours). As a result, cells can potentially slip past each other while maintaining overall tissue cohesion. Such neighbor changes could cause epithelial tissue to behave as a viscous fluid on long time scales rather than an elastic solid, as it does during convergent extension~\cite{Wang13541}. Vertex models~\cite{doi:10.1080/13642810108205772,FLETCHER20142291,pmid18082406,C3SM52893F,Bi_nature_physics,PhysRevX.6.021011,D0SM01575J,Hufnagel3835,Rauzi_et_al,LANDSBERG20091950}  have provided a useful framework for how tissues can switch between solid and fluid behavior~\cite{Bi_nature_physics, PhysRevX.6.021011}, and have had remarkable success in describing experimental results~\cite{Atia_et_al, pmid26237129, Wang13541, PhysRevX.9.011029, 10.1016/j.bpj.2019.09.027a}. These models make the central assumption that internal forces within a tissue are approximately balanced on time scales intermediate between ms and hours, and have successfully described phenomena such as pattern formation, cell dynamics, and cell movement during tissue development ~\cite{staple_mechanics_2010}. Force balance is captured by minimizing an energy that depends on cell shapes. In the standard vertex model, energy barriers are lower when cells have high aspect ratios, so higher/lower cell aspect ratios correspond to fluid/solid behavior. 

During dorsal closure, significant changes in AS cell shapes are observed. According to the standard vertex model, the observed high values of mean cell shape aspect ratio should render the tissue fluid~\cite{C3SM52893F, Bi_nature_physics, PhysRevX.6.021011}. Nonetheless, there is considerable experimental evidence that the AS remains \emph{solid} during dorsal closure with \emph{no} neighbor exchanges~\cite{10.1073/pnas.1018652108,Ma_2009,machado_emergent_2015}. We have examined individual junction lengths using live embryo imaging in an extensive data set comprising 10s of embryos, each with 100s of cells and, in agreement with the literature, did not find any vanishing junctions, and hence any neighbor exchanges, except when cells left the AS (cell ingression). 

Vertex models might simply fail to describe tissue mechanics at this stage of development. The success of vertex models in describing many other tissues, however, begs the question: can the models be tweaked to capture tissue mechanics of the AS during dorsal closure, and could this point to an important physiological control mechanism? Here we introduce a minimal extension to the standard vertex model that quantitatively captures results from comprehensive experimental datasets obtained from time-lapse microscopy recordings.

\section{Modeling and experimental analysis}
\begin{figure*}[ht]
\center
\includegraphics[width=\textwidth]{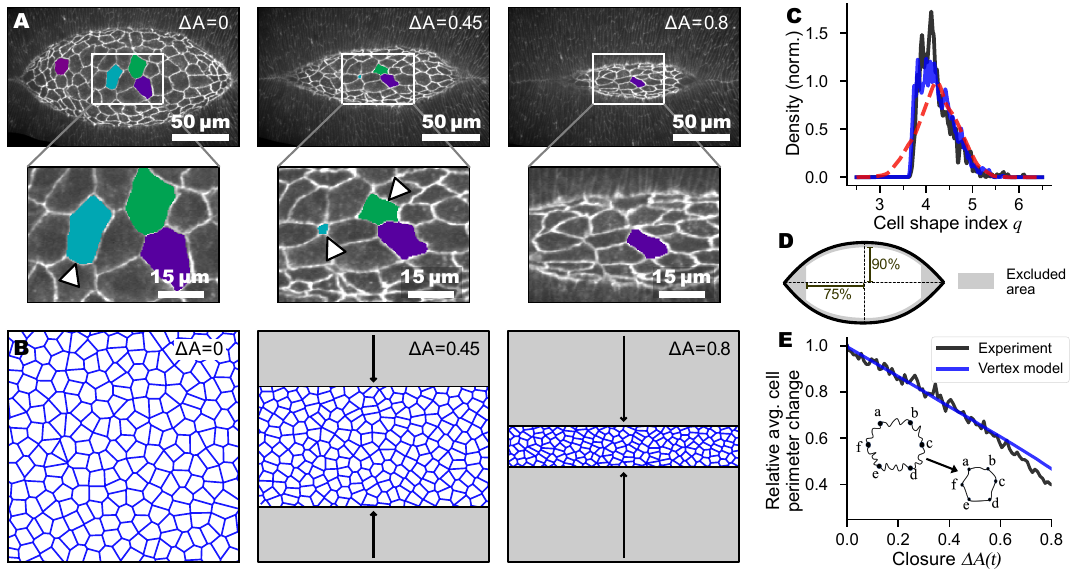}
\caption{\textbf{Experiment and vertex model for dorsal closure}. (\textbf{A}) The geometry of the dorsal hole during  early (left), middle (center), and late (right) dorsal closure. Enlargements show tissue with selected cells, several of which ingress (highlighted by triangles). (\textbf{B}) We model the dorsal closure process as a quasistatic uni-axial deformation. The geometry of the model is shown at the beginning (left), in the middle (center, at $45\,\%$ closure), and towards the end (right, $80\,\%$ closure) of the process. $\Delta A(t) = \frac{A_0-A(t)}{A_0}$ is the fractional change in total AS area of the closure process, where $A_0$ is the AS area at the onset of dorsal closure. (\textbf{C}) An initial normal distribution of the preferred shape index of the model tissue (dashed red) with the standard deviation adjusted to be 0.45, leads to a distribution of the actual shape index after minimization (solid blue) that is in excellent agreement with the distribution of the experimentally observed shape index (solid black) at the beginning of dorsal closure. (\textbf{D}) Sketch of AS tissue regions included in model comparison (white center), with edge regions excluded (gray regions). (\textbf{E}) In the model, we reduce the preferred cell perimeter at a linear rate (blue) to capture the experimentally observed decrease of junction lengths (black). For comparison, we normalize the average perimeter by its value at the onset of the process. Inset: schematic representation of the reduction of cellular junction length and apical area during dorsal closure. }
\label{fig:main_1}
\end{figure*}

Our starting point is a standard two-dimensional cellular vertex model 
~\cite{brodland_differential_2002,doi:10.1080/13642810108205772,farhadifar_influence_2007, FLETCHER20142291,pmid18082406,C3SM52893F,Bi_nature_physics,D0SM01575J,Hufnagel3835,Rauzi_et_al,LANDSBERG20091950, PhysRevX.6.021011, C3SM52893F} (short introduction to vertex models in SI section\,\ref{SI_intro_vertex_models}). The AS is represented as a single-layer sheet of polygonal cells that tile the entire area, as described below. In our model, we approximate the shape of the AS tissue (Fig.\,\ref{fig:main_1}A) with a rectangle whose long axis corresponds to the anterior-posterior axis of the embryo (Fig.\,\ref{fig:main_1}B, details see SI section \ref{SI_model_details}). During simulated dorsal closure, the positions of the vertices are continually adjusted to maintain the mechanical energy of the tissue at a minimum, or equivalently, to balance the forces exerted on each vertex. The mechanical energy of the standard vertex model is defined as 
\begin{equation}
\label{eq:vertex}
E = \sum^{N}_{i=1} \frac{1}{2}k_a (a_i - a_{0})^2 + \frac{1}{2}k_p (p_i - p_{0,i})^2,
\end{equation}
where $N$ is the total number of cells, $p_i$ and $a_i$ are the actual cell perimeters and areas, $p_{0,i}$ and $a_{0}$ are the preferred cell perimeters and area, and $k_p$ and $k_a$ represent the perimeter and area elastic moduli of the cells, respectively. The first term penalizes apical area changes away from a preferred value, and can arise from cell height changes as well as active contractions in the medio-apical actin network at constant or near constant volume. The second term combines the effects of actomyosin cortex contractility with cell-cell adhesion, where $p_{0,i}$ is the effective preferred cell perimeter~\cite{Bi_nature_physics}. For simplicity, we chose $k_a=k_p=1$ for all cells. 

We used time-lapse confocal microscopy to image the entire dorsal closure process in E-cadherin-GFP embryos. We then used our custom machine-learning-based cell segmentation and tracking algorithm to create time series of cell centroid position, area, perimeter, aspect ratio, and individual junction contour lengths for every cell in the AS \cite{deeptissue}.
At the onset of closure we find that cells in the AS exhibit considerable variability of the cell shape index $ q_i = p_i/\sqrt a_i $ (Fig.\,~\ref{fig:main_1}C). In the model, we therefore introduce initial polydispersity in the cell shape index through a normal distribution of preferred cell perimeters $p_{0,i}$.  We fix the preferred cell area and use it to set our units so that $a_{0}=1$ for all cells, following Ref.~\cite{PhysRevLett.123.058101}.
The distribution of actual shape index $q_i$ after minimizing the mechanical energy in the model is in excellent agreement with the experiments (Fig.\,~\ref{fig:main_1}C). 

During a substantial part of closure (Fig.\,~\ref{fig:main_1}A), the leading edges of the two flanking epithelial sheets approach the dorsal mid-line at a roughly constant rate~\cite{doi:10.1126/science.1079552}. To mimic these dynamics, we linearly decreased the vertical height of the rectangle representing the AS (Fig.\,~\ref{fig:main_1}B) by 0.125 \% of the initial height at every step while holding the width fixed. We enforce force balance, minimizing the mechanical energy after each deformation step. We used periodic boundary conditions throughout.

Since closure rates varied from embryo to embryo, we measured progress during closure not in terms of \emph{time}, but in terms of fractional change of the total area of the exposed AS (i.e. the dorsal opening), $\Delta A(t) = \frac{A_0-A(t)}{A_0}$, where $A_0$ is a reference area of the AS early during closure. In many prior studies~\cite{PERALTA20072583,wells_complete_2014}, the height of the AS has been used as a descriptor of closure progress. In Fig.\,\ref{SI_Fig_ARea_vs_height} we demonstrate that both height and area of the AS decreased monotonically and approximately linearly with time, validating our use of $\Delta A(t)$ to mark the progression of closure. We began the analysis of each embryo at $A_0 = 11,000\,\mu\text{m}^2$, so that we could average over multiple embryos. To exclude complex tissue boundary effects, we excluded cells at the AS borders and the regions at the canthi in the comparison between model and experiment (Fig.\,~\ref{fig:main_1}D).

AS cells reduce their perimeter (inset Fig.\,\ref{fig:main_1}E) \cite{pmid30458138} during the closure process by removing a portion of junction material and membranes through endocytosis, while maintaining junction integrity \cite{pmid21516109,Mateus}. The average perimeter shrinks at a constant rate in the experiments (Fig.\,~\ref{fig:main_1}E). We therefore assume in the model that the \textit{preferred} perimeter $p_{0,i}$ of each cell decreases linearly with $\Delta A(t)$ at the same rate (details see SI section \ref{SI_model_details}).  Note that we do not change the preferred area per cell, $a_0=1$. For a more realistic model we could change the preferred area in proportion to the total area of the AS as it shrinks, but that would change only the pressure, and would have no effect on the rigidity transition~\cite{yang_correlating_2017, teomy_confluent_2018,arzash_tuning_2023}. 

During dorsal closure, $\sim 10\,\%$ of AS cells ingress into the interior of the embryo~\cite{pmid10769037,pmid22404919,doi:10.1146/annurev-cellbio-111315-125357} (additional cells ingress at the canthi and adjacent to the lateral epidermis). In the model, we removed cells randomly at the experimentally measured rate (see details in SI) so that roughly 10\,\% of the AS cells disappeared over the course of dorsal closure. 

For further details of the model and the experiments, see Materials and Methods and Supplemental Information.
\section{Results}

\begin{figure*}[h]
\center
\includegraphics[width=\textwidth]{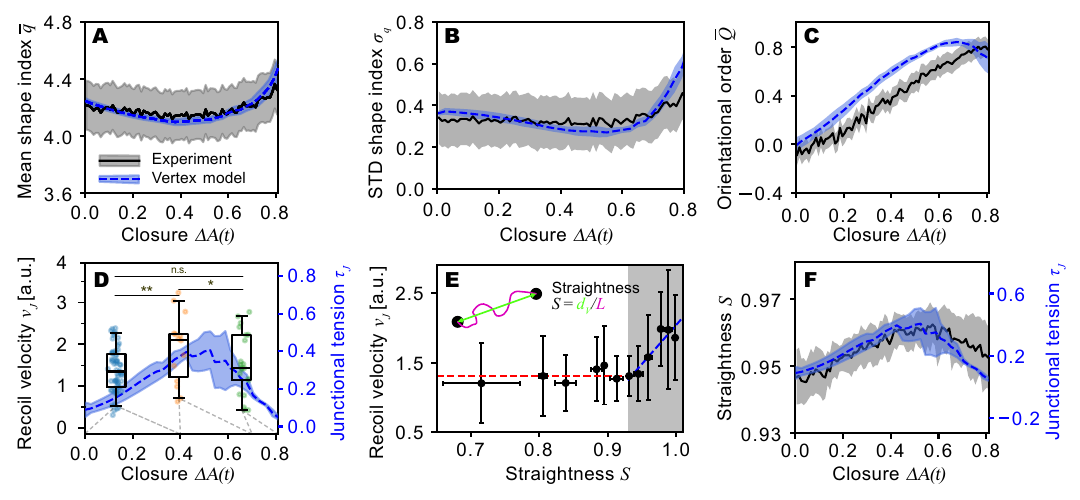}
\caption{\textbf{Results from experiment (black solid) and model (blue dashed).} (\textbf{A}) A comparison of average shape index $\bar q$ as a function of $\Delta A(t) = \frac{A_0-A(t)}{A_0}$. Here $A_0$ is the AS area at the onset of dorsal closure and $A(t)$ is the area as it shrinks during dorsal closure, so that $\Delta A(t)=0$ at onset. Lines show the mean, and shaded regions indicate the standard deviation among 12 embryos (experiment) or 10 different initial configurations (model). (\textbf{B}) Comparison of cell to cell standard deviation of the shape index ($\sigma_q$) during dorsal closure. (\textbf{C}) Orientational order parameter ($\bar{Q}$) of the cells during dorsal closure. (\textbf{D}) Experimental initial junction recoil velocity (left y-axis) of the vertices after performing laser ablation of the junction, and predicted average cellular cortical tension ($\bar{\tau_J}$)(right y-axis) of the model during dorsal closure. The boxplots represent data across three intervals of $\Delta A$ ($\Delta A<0.4$, $0.4 \leq \Delta A < 0.7$, $\Delta A \geq 0.7$). Whiskers extend to the 5th and 95th percentiles, while the boxes delineate the interquartile range, and the horizontal lines within the boxes indicate the median values. An ANOVA followed by a post-hoc Tukey's HSD test was conducted to assess statistical significance (*: $p<0.1$, **: $p<0.05$). We performed and evaluated cuts of $N=97$ junctions. (\textbf{E}) Average initial recoil velocity of vertices after laser cutting as a function of junction straightness (ratio of the inter-vertex distance ($d_v$) to the junction length ($L$), see inset) immediately before cutting. Junction recoil velocity is independent of junction straightness (fitted with the red dashed line) until $S = d_v/L \gtrsim 0.93$. The crossover point at $d_v/L \approx 0.93$ marks the intersection of the red and blue dashed lines; the latter fits the data points in the gray-shaded region, indicating that the recoil velocity increases strongly and approximately linearly with junction straightness in this regime. (\textbf{F}) Comparison of experimental junction straightness (left y-axis) and model cellular junction tension (right y-axis) during dorsal closure.}
\label{fig:main_2}
\end{figure*}

We tracked the following quantities during dorsal closure, in model and experiments: mean cell shape index $\bar q = \big\langle p_i/\sqrt{a_i} \big\rangle$, mean aspect ratio $\bar {\alpha}$ (see SI section \ref{SI_shape_index}), orientational order parameter $\bar{Q} = \langle \cos(2\theta) \rangle$ \cite{SF9710500016} (see SI section \ref{SI_orientational_order_parameter}) characterizing the degree of cellular alignment (where $\theta$ is the angle between the major axis of each cell and the anterior-posterior axis, $\bar{Q}=0$ for randomly aligned cells and $\bar{Q}=1$ for cells perfectly aligned with the AP axis), standard deviation of cell shape index $\sigma_q$ and standard deviation of the aspect ratio $\sigma_\alpha$.

We compare experimental data and simulations without any parameter modifications, adjustments, or rescaling with time. Considering the simplicity of the model, the agreement is remarkably good, both for cell shape and cell shape variability (Fig.\,\ref{fig:main_2}A,B) as well as cellular alignment (Fig.\,\ref{fig:main_2}C). The mean and standard deviation of cell aspect ratio agree equally well (SI\,\ref{SI_AR_STD_mean} and Fig.\,\ref{SI_Fig_AR}). As expected, the error bars (shaded region) in the experimental data, which represent variations between different embryos, are significantly larger than those in the simulation data, which represent only variations between initial configurations based on a single distribution of cell shape indices from the distribution measured over all embryos (Fig.~\ref{fig:main_1}C). In the experiments, there is intrinsic embryo-to-embryo variability that we did not include in our model for simplicity.

In experiment and model, the mean shape index $\bar q$ initially decreases, reaches a minimum at $\Delta A \approx 0.55$ and then increases (Fig.\,\ref{fig:main_2}A). In the model, this behavior arises from two competing effects. (i) Decreasing preferred mean perimeter (Fig.\,~\ref{fig:main_1}E) implies a decreasing preferred mean shape index, $\bar q_0 = \langle p_{0,i}/\sqrt{ a_{0,i}} \rangle$. According to Eq.~\ref{eq:vertex}, this tends to drag down $\bar q$, causing the decrease up to $\Delta A \approx 0.55$. (ii) As dorsal closure progresses, the overall shape of the tissue becomes more and more anisotropic (Fig.\,\ref{fig:main_1}A,B), increasing $\bar q$. This effect eventually dominates for $\Delta A \gtrsim 0.55$. This competition between decreasing $\bar q_0$ and increasing anisotropy is also reflected in the width of the $q$-distribution, $\sigma_q$ (Fig.\,~\ref{fig:main_2}B). In the model, the standard deviation of the preferred shape index, $\sigma_{q,0}$, is fixed, but cell to cell variations of the energy $E$ due to $\sigma_{q,0}$ increase with decreasing $\bar q_0$, leading to a narrowing of the distribution, or a decrease in $\sigma_q$.  On the other hand, vertical shrinking of the system late in the closure process leads to greater $\sigma_q$ (Fig.\,\ref{fig:main_2}B). The increasing anisotropy during closure leads to greater alignment of cells along the anterior-posterior axis, reflected in an increased orientational order parameter $\bar Q$ (Fig.\,~\ref{fig:main_2}C).

A strength of the vertex model is that it predicts not only cell shape and orientation distributions but also mechanical cell-level properties of the AS, including, for example, the  cell junction tension $\tau_J$, defined as \cite{PhysRevX.9.011029,doi:10.1073/pnas.1705921114}
\begin{equation}
\tau_J = k_p (p_i-p_{0,i}) + k_p (p_j-p_{0,j}),
\label{eq:junctionaltension}
\end{equation}
where $i,j$ denote cells that share a given junction. Relative values of junction tension can be estimated experimentally from the initial recoil velocity $v_r$ when a junction is severed using laser ablation \cite{pmid26389664,pmid21068726,pmid18978783,pmid19879198,pmid18082406}. Our model predicts that the average junction tension  $\bar \tau_J$ rises until the fractional area of the AS reaches $\Delta A(t) \approx 0.55$, and then decreases as dorsal closure continues (Fig.\,\ref{fig:main_2}D). To test this prediction, we conducted laser cutting experiment at different stages of closure. We find that the recoil velocity changed in a non-monotonic manner (Fig.\,~\ref{fig:main_2}D), as the model predicts.

An alternative way to estimate junction tension from imaging data of unperturbed embryos is to analyze the straightness of junctions. A wiggly junction would be expected to be free of tension, whereas a straight junction should support tension. We define junction straightness as $S=d_v/L$ (Fig.\,~\ref{fig:main_2}E, inset), where $d_v$ is the distance between vertices for a given junction and $L$ is the contour length of the junction. We examined the relation between $S$ and the initial recoil velocity upon cutting, $v_r$, and observed that $v_r$ is independent of $S$ for $S\lesssim 0.93$, but then rises linearly with increasing $S$ above this threshold (Fig.\,~\ref{fig:main_2}E). It is reasonable to assume that junction straightness $S$ is proportional to the tension $\tau_J$ predicted by our model. This is verified in Fig.\,~\ref{fig:main_2}F, showing the same non-monotonicity for both quantities with peaks occurring at $\Delta A(t) \approx 0.55$.

Why is the junction tension non-monotonic? In vertex models, junction tension and cell stiffness are related to cell shape index \cite{Atia_et_al,Bi_nature_physics, doi:10.1073/pnas.1705921114,pmid26237129}. According to Eq.\,~\ref{eq:junctionaltension}, junction tension is given by the difference between cell preferred perimeter $p_{0.i}$ and the actual perimeter $p_i$ for the two cells sharing a given junction. Below $\Delta A = 0.55$, $p_i-p_{0,i}$ increases with $\Delta A$, leading to an increase of $\bar \tau_J$. For $\Delta A \geq 0.55$, $p_i-p_{0.i}$ decreases, leading to a decrease of $\bar \tau_J$.

A striking result of the standard vertex model (Eq.~\ref{eq:vertex}) is the prediction of a transition from solid to fluid behavior as the average shape index increases above $\bar q_c=3.81$~\cite{Bi_nature_physics}, in excellent agreement with a number of experiments in various epithelial tissue models \cite{Atia_et_al, lawson-keister_jamming_2021,mongera_fluid--solid_2018,petridou_rigidity_2021}. Inspection of Fig.\,~\ref{fig:main_2}A shows that $\bar q > 3.81$ during the entire process of dorsal closure, suggesting that the AS should be fluid. However, the complete absence of T1 events (cell neighbor changes) shows conclusively that the AS is not fluid but solid.

Which of the extensions of the standard vertex model (Eq.~\ref{eq:vertex}) that we have incorporated in our model are responsible for the solid nature of the AS? It is known that cellular shape heterogeneity~\cite{PhysRevLett.123.058101} and orientational ordering~\cite{Wang13541} both enhance rigidity in vertex models (detailed analysis of orientational alignment see SI section \ref{SI_anisotropy_alignment}). In our case, cellular heterogeneity remains essentially constant during dorsal closure, but orientational ordering increases due to uniaxial deformation. Isotropic deformation (SI\, section \ref{SI_isotropic_constriction}), in contrast, does not lead to orientational order (Fig.\,\ref{SI_Fig_isotropic}B), as one might expect. The incorporation of uniaxial deformation is important since trends in $\bar q$, $\sigma_q$ and junction tension (Fig.\,\ref{SI_Fig_isotropic}A,D,C) with closure fail to agree with experimental results if we apply isotropic deformation. However, we find that our model predicts solid behavior even for isotropic deformation, showing that uniaxial deformation is not needed for this aspect. We also find that cell ingression at the levels seen experimentally has almost no effect on the solid  behavior in the model. This leaves the progressive decrease of preferred cell perimeter as crucial for maintaining solid response.

For the tissue to behave as a solid, non-zero tension cell junctions must form continuous paths that extend across the entire system in all directions~\cite{PhysRevLett.123.058101,Petridou2021-ow}--in other words, they must \emph{percolate}. Percolation requires the fraction of junctions with non-zero tension, $f_r$ to be larger than a critical fraction $f_c$. The rigidity transition can be driven either by altering  $f_r$ or  $f_c$, or both. Note that for random Voronoi tessellations in a square system, $f_c \approx 0.66$ \cite{doi:10.1080/00018737100101261,PhysRevE.60.6361,PhysRevE.80.041101}. The topology of such networks is similar to that of the standard vertex model (Eq. \ref{eq:vertex}), so that $f_c \approx 0.66$ can be taken as a reasonable approximation. We show in the Supplemental Information section \ref{SI_percolation} that $f_c$ remains fixed during dorsal closure, despite uniaxial deformation of the AS. Interestingly, the tissue would fluidize without the progressive decrease of preferred cell perimeter (Fig.\,\ref{fig:SI_percolation}B).

\begin{figure}[h]
\centering
\includegraphics[width=0.4\textwidth]{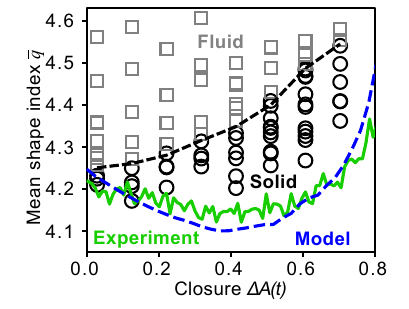}
\caption{\textbf{Phase diagram of amnioserosa during dorsal closure.} Phase diagram in $\bar q$ vs.~$\Delta A$ space, with solid (open black circles) and fluid (open gray squares) states as shown. The solid-fluid transition is marked by the black dashed line. Both experiment (green dashed line) and model (blue dashed line) are within the solid phase throughout dorsal closure.}
  \label{fig:main_3}
\end{figure} 

Fig.\,~\ref{fig:main_3} summarizes our results in the form of a phase diagram of our model, obtained by evaluating the fraction of rigid junctions $f_r$ at discrete values of mean cell shape index and $\Delta A$. The phase boundary corresponds to the percolation transition of nonzero-tension junctions, $f_r=f_c \approx 0.66$. The system always remains in the solid phase, consistent with the experimental observations (blue dashed line). As explained earlier, $\bar q$ initially decreases because the decreasing preferred perimeter pulls actual cell perimeters down. Eventually, however, the elongation of cells due to uniaxial deformation overcomes this effect, causing $\bar q$ to increase.

\section{Discussion} 
We find experimentally that the AS remains in a solid phase (\textit{i.e.} with no cell neighbor exchanges) during dorsal closure. One might not be surprised since cells adhere to each other. It is important to realize, though, that cadherins have rapid on-off kinetics and the actin cortex has rapid turnover on the time scale of dorsal closure. As a result, adhesion cannot necessarily prevent cell neighbor switching; it merely guarantees tissue cohesion. Indeed, the standard vertex model predicts that when cells are highly elongated, the barriers to neighbor switching should be low and the tissue should be fluid~\cite{PhysRevX.6.021011,pmid26237129,Atia_et_al,Wang13541,10.1073/pnas.1917853118}. from the AS cell shapes that barriers should be low and that the AS should be a fluid. Our simple extension of the standard vertex model not only predicts that the tissue should be in the solid phase but also faithfully reproduces a wide range of characteristics of an extensive set of experimental dorsal closure data: cell shape and  orientational order, and junction tension, which we inferred passively from image data due to the linear relationship between junction straightness and initial recoil velocity in laser cutting experiments. 

Our model achieves this good agreement with only two parameters that are directly derived from experiments. We find that shape polydispersity and active shrinking of the preferred cell perimeters are the two critical factors that enable the tissue to remain solid in spite of extensive cellular and tissue shape changes. These results imply that the solid character of the AS originates from active processes that regulate cell perimeter, including junction complexes and the components of the cell cortex.

This finding raises two questions for future research. First, \emph{how} is the removal of junction material specifically regulated in cells? Second, \emph{why} might it be important for the AS to remain in a solid phase? Perhaps solid behavior during dorsal closure is simply a holdover from the preceding developmental stage of germ band retraction \cite{Lan_2015}. Laser ablation experiments \cite{lynch_cellular_2013} suggest that the AS plays an important assistive role in uncurling of the germ band by exerting anisotropic tension on it. Such anisotropic stress requires the AS to be a solid, not fluid. An interesting future direction for experimental and vertex model studies is to establish whether the AS is solid throughout germ band retraction as well as dorsal closure.

Our results show that vertex models are more broadly applicable than previously thought. Despite the many complex active processes that occur during dorsal closure, we find that only one of them--the active shrinking of a normally-fixed parameter, namely the preferred perimeter--is needed in order to quantitatively describe our experimental observations.  Similar variation of normally-constant parameters has been shown to allow other systems to develop complex responses not ordinarily observed in passive non-living systems. These include negative Poisson ratios \cite{PhysRevLett.114.225501,hexner_role_2018} and allostery \cite{rocks_designing_2017} in mechanical networks, greatly enhanced stability in particle packings \cite{hagh_transient_2022}, and the ability to classify data and perform linear regression in mechanical and flow networks \cite{PhysRevX.11.021045} as well as laboratory electrical networks \cite{dillavouPRApplied2022}. More generally, the mechanical behavior of epithelial tissues during development is extraordinary when viewed through the lens of ordinary passive materials.  It remains to be seen how much of that behavior can be understood using "adaptive vertex models"~\cite{arzash_tuning_2023} within a framework that replaces ordinarily fixed physical parameters with degrees of freedom that vary with time.

\matmethods{}
\showmatmethods{}
Flies were maintained using standard methods, and embryos were collected and prepared for imaging and laser surgery as previously described \cite{z_drosophila_1991,KIEHART200687, wells_complete_2014, goldstein_chapter_1994}. Cell junctions were labeled via ubiquitous expression of DE-cadherin-GFP \cite{oda_real-time_2001}. Images were captured using Micro-Manager 2.0 software (Open Imaging) to operate a Zeiss Axiovert 200 M microscope outfitted with a Yokogawa CSU-W1 spinning disk confocal head (Solamere Technology Group), a Hamamatsu Orca Fusion BT camera, and a Zeiss 40X LD LCI PlanApochromat 1.2\,NA multi-immersion objective (glycerin). Due to the embryo's curvature, multiple z planes were imaged for each embryo at each time point to observe the dorsal opening. We recorded stacks with eight z-slices with 1\,µm step size every 15\,s throughout the closure duration, with a 100\,ms exposure per slice. 

Two-dimensional projections of the AS tissue were created from 3D stacks using DeepProjection \cite{haertter_deepprojection_2022}. A custom Python algorithm was used to segment and track individual AS cells throughout dorsal closure \cite{deeptissue}: Briefly, binary masks of the AS cell boundaries and the amnioserosa tissue boundary (leading edge) were first predicted from microscopy movies using deep learning trained with expert-annotated dorsal closure specific data \cite{10.1038/s41592-018-0261-2}. Second, individual AS cells were segmented and tracked throughout the process using the watershed segmentation algorithm with propagated segmentation seeds from previous frames. Finally, for each cell, area, perimeter, aspect ratio and orientation in relation to the AS anterior-posterior axis were quantified over time. Based on the binary mask of the leading edge, we segmented the dorsal hole/AS shape, fitted an ellipse to it at each time point, and located the centroid position of each cell with respect to the long and short axis of the ellipse. This allowed us to precisely identify cells in the amnioserosa center (within 75\,\% of the semi-major axis and 90\,\% of the semi-minor axis), and exclude peripheral cells from comparisons between model and experiment. The straightness $S$ of cell-cell junctions was quantified by segmenting the contour and end-to-end lengths of individual junctions using a graph-based algorithm~\cite{deeptissue}. 

Laser surgery was performed on a Zeiss Axio Imager M2m microscope equipped with a Yokogawa CSU-10 spinning disk confocal head (Perkin Elmer), a Hamamatsu EM-CCD camera and a Zeiss 40X, 1.2\,NA water immersion objective. Micro-Manager 1.4.22 software (Open Imaging) controlled the microscope, the Nd:YAG UV laser minilite II (Continuum, 355\,nm, 4\,mJ, 1.0\,MW peak power, 3–5\,ns pulse duration, 10\,Hz, \cite{celis_chapter_2006}) and a steering  mirror for laser incisions. In each embryo (N = 48), 1 to 2 cuts of approx. 5\,µm length with a laser setting at 1.4\,µJ were performed in the bulk of the AS at different stages of closure \cite{KIEHART200687, 10.2976/1.2955565, 10.1091/mbc.e14-07-1190} (Fig.\,\ref{SI_Fig_cut}A,B). The response of the AS was recorded prior to ($\sim$ 20 frames), during ($\sim$ 4 frames) and after ($\sim$ 576 frames) the cut at a frame rate of 5\,Hz. The junction straightness $S$ of each cut junction was manually quantified prior to the cut by manually tracing junction end-to-end length and junction contour length using ImageJ. Then, to analyze the initial recoil velocity, the motion of the vertices adjacent to the cut junction was followed in a kymograph perpendicular to the cut (line thickness 2\,µm, Fig.\,\ref{SI_Fig_cut}A-C). On the basis of the kymograph, the distance $d(t)$ between the two vertices of the severed junction was quantified manually over time using ImageJ. A double exponential function $a_0 \exp(b_0t) + c_0 \exp(d_0t) +e_0$ was fitted to $d(t)$ (Fig.\,\ref{SI_Fig_cut}D). The initial slope of this function at $t \sim 0$ corresponds to the initial recoil velocity $v_r$.

For the vertex model, we used the open-source CellGPU code \cite{SUSSMAN2017400}. Analysis and illustration of model and experiment data was performed with custom Python scripts. Simulation code will be published on GitHub upon publication. Data associated with this study are available upon request.

\acknow{We thank M. L. Manning and S. R. Nagel for instructive discussions. This project was supported by NIH through Awards R35GM127059 (DPK) and 1-U01-CA-254886-01 (IT), NSF-DMR-MT-2005749 (IT, AJL) and by the Simons Foundation through Investigator Award \#327939 (AJL). AJL thanks CCB at the Flatiron Institute, as well as the Isaac Newton Institute for Mathematical Sciences under the program "New Statistical Physics in Living Matter" (EPSRC grant EP/R014601/1), for support and hospitality while a portion of this research was carried out.}

\showacknow{} 


\bibliography{Dorsal_Closure}

\begin{thebibliography}{10}

\bibitem{harden_signaling_2002}
N Harden, Signaling pathways directing the movement and fusion of epithelial sheets: lessons from dorsal closure in {Drosophila}.
\newblock {\em\protect\JournalTitle{Differentiation; Research in Biological Diversity}} \textbf{70}, 181--203 (2002).

\bibitem{hayes_drosophila_2017}
P Hayes, J Solon, Drosophila dorsal closure: {An} orchestra of forces to zip shut the embryo.
\newblock {\em\protect\JournalTitle{Mechanisms of Development}} \textbf{144}, 2--10 (2017).

\bibitem{doi:10.1146/annurev-cellbio-111315-125357}
DP Kiehart, JM Crawford, A Aristotelous, S Venakides, GS Edwards, Cell sheet morphogenesis: Dorsal closure in drosophila melanogaster as a model system.
\newblock {\em\protect\JournalTitle{Annual Review of Cell and Developmental Biology}} \textbf{33}, 169--202 (2017) PMID: 28992442.

\bibitem{aristotelous_mathematical_2018}
A Aristotelous, J Crawford, G Edwards, D Kiehart, S Venakides, Mathematical {Models} of {Dorsal} {Closure}.
\newblock {\em\protect\JournalTitle{Progress in biophysics and molecular biology}} \textbf{137}, 111--131 (2018).

\bibitem{young_morphogenesis_1993}
PE Young, AM Richman, AS Ketchum, DP Kiehart, Morphogenesis in {Drosophila} requires nonmuscle myosin heavy chain function.
\newblock {\em\protect\JournalTitle{Genes \& Development}} \textbf{7}, 29--41 (1993) Company: Cold Spring Harbor Laboratory Press Distributor: Cold Spring Harbor Laboratory Press Institution: Cold Spring Harbor Laboratory Press Label: Cold Spring Harbor Laboratory Press Publisher: Cold Spring Harbor Lab.

\bibitem{kiehart_multiple_2000}
DP Kiehart, CG Galbraith, KA Edwards, WL Rickoll, RA Montague, Multiple {Forces} {Contribute} to {Cell} {Sheet} {Morphogenesis} for {Dorsal} {Closure} in {Drosophila}.
\newblock {\em\protect\JournalTitle{Journal of Cell Biology}} \textbf{149}, 471--490 (2000) \_eprint: https://rupress.org/jcb/article-pdf/149/2/471/1430839/9910093.pdf.

\bibitem{doi:10.1126/science.1079552}
MS Hutson, et~al., Forces for morphogenesis investigated with laser microsurgery and quantitative modeling.
\newblock {\em\protect\JournalTitle{Science}} \textbf{300}, 145--149 (2003).

\bibitem{franke_nonmuscle_2005}
JD Franke, RA Montague, DP Kiehart, Nonmuscle {Myosin} {II} {Generates} {Forces} that {Transmit} {Tension} and {Drive} {Contraction} in {Multiple} {Tissues} during {Dorsal} {Closure}.
\newblock {\em\protect\JournalTitle{Current Biology}} \textbf{15}, 2208--2221 (2005).

\bibitem{pmid30399339}
D Pinheiro, Y Bellaiche, {{M}echanical {F}orce-{D}riven {A}dherens {J}unction {R}emodeling and {E}pithelial {D}ynamics}.
\newblock {\em\protect\JournalTitle{Dev Cell}} \textbf{47}, 391 (2018).

\bibitem{pmid31251912}
E Hannezo, CP Heisenberg, {{M}echanochemical {F}eedback {L}oops in {D}evelopment and {D}isease}.
\newblock {\em\protect\JournalTitle{Cell}} \textbf{178}, 12--25 (2019).

\bibitem{Wang13541}
X Wang, et~al., Anisotropy links cell shapes to tissue flow during convergent extension.
\newblock {\em\protect\JournalTitle{Proceedings of the National Academy of Sciences}} \textbf{117}, 13541--13551 (2020).

\bibitem{doi:10.1080/13642810108205772}
T Nagai, H Honda, A dynamic cell model for the formation of epithelial tissues.
\newblock {\em\protect\JournalTitle{Philosophical Magazine B}} \textbf{81}, 699--719 (2001).

\bibitem{FLETCHER20142291}
AG Fletcher, M Osterfield, RE Baker, SY Shvartsman, Vertex models of epithelial morphogenesis.
\newblock {\em\protect\JournalTitle{Biophysical Journal}} \textbf{106}, 2291--2304 (2014).

\bibitem{pmid18082406}
R Farhadifar, JC Röper, B Aigouy, S Eaton, F Jülicher, {{T}he influence of cell mechanics, cell-cell interactions, and proliferation on epithelial packing}.
\newblock {\em\protect\JournalTitle{Curr Biol}} \textbf{17}, 2095--2104 (2007).

\bibitem{C3SM52893F}
D Bi, JH Lopez, JM Schwarz, ML Manning, Energy barriers and cell migration in densely packed tissues.
\newblock {\em\protect\JournalTitle{Soft Matter}} \textbf{10}, 1885--1890 (2014).

\bibitem{Bi_nature_physics}
D Bi, JH Lopez, JM Schwarz, ML Manning, A density-independent rigidity transition in biological tissues.
\newblock {\em\protect\JournalTitle{Nature Physics}} \textbf{11}, 1074--1079 (2015).

\bibitem{PhysRevX.6.021011}
D Bi, X Yang, MC Marchetti, ML Manning, Motility-driven glass and jamming transitions in biological tissues.
\newblock {\em\protect\JournalTitle{Phys. Rev. X}} \textbf{6}, 021011 (2016).

\bibitem{D0SM01575J}
I Tah, TA Sharp, AJ Liu, DM Sussman, Quantifying the link between local structure and cellular rearrangements using information in models of biological tissues.
\newblock {\em\protect\JournalTitle{Soft Matter}} pp.~-- (2021).

\bibitem{Hufnagel3835}
L Hufnagel, AA Teleman, H Rouault, SM Cohen, BI Shraiman, On the mechanism of wing size determination in fly development.
\newblock {\em\protect\JournalTitle{Proceedings of the National Academy of Sciences}} \textbf{104}, 3835--3840 (2007).

\bibitem{Rauzi_et_al}
M Rauzi, P Verant, T Lecuit, PF Lenne, Nature and anisotropy of cortical forces orienting drosophila tissue morphogenesis.
\newblock {\em\protect\JournalTitle{Nature Cell Biology}} \textbf{10}, 1401--1410 (2008).

\bibitem{LANDSBERG20091950}
KP Landsberg, et~al., Increased cell bond tension governs cell sorting at the drosophila anteroposterior compartment boundary.
\newblock {\em\protect\JournalTitle{Current Biology}} \textbf{19}, 1950--1955 (2009).

\bibitem{Atia_et_al}
L Atia, et~al., Geometric constraints during epithelial jamming.
\newblock {\em\protect\JournalTitle{Nature Physics}} \textbf{14}, 613--620 (2018).

\bibitem{pmid26237129}
JA Park, et~al., {{U}njamming and cell shape in the asthmatic airway epithelium}.
\newblock {\em\protect\JournalTitle{Nat Mater}} \textbf{14}, 1040--1048 (2015).

\bibitem{PhysRevX.9.011029}
L Yan, D Bi, Multicellular rosettes drive fluid-solid transition in epithelial tissues.
\newblock {\em\protect\JournalTitle{Phys. Rev. X}} \textbf{9}, 011029 (2019).

\bibitem{10.1016/j.bpj.2019.09.027a}
MF Staddon, KE Cavanaugh, EM Munro, ML Gardel, S Banerjee, Mechanosensitive {{Junction Remodeling Promotes Robust Epithelial Morphogenesis}}.
\newblock {\em\protect\JournalTitle{Biophysical Journal}} \textbf{117}, 1739--1750 (2019).

\bibitem{staple_mechanics_2010}
DB Staple, et~al., Mechanics and remodelling of cell packings in epithelia.
\newblock {\em\protect\JournalTitle{The European Physical Journal E}} \textbf{33}, 117--127 (2010).

\bibitem{10.1073/pnas.1018652108}
C Meghana, et~al., Integrin adhesion drives the emergent polarization of active cytoskeletal stresses to pattern cell delamination.
\newblock {\em\protect\JournalTitle{Proceedings of the National Academy of Sciences}} \textbf{108}, 9107--9112 (2011).

\bibitem{Ma_2009}
X Ma, HE Lynch, PC Scully, MS Hutson, Probing embryonic tissue mechanics with laser hole drilling.
\newblock {\em\protect\JournalTitle{Physical Biology}} \textbf{6}, 036004 (2009).

\bibitem{machado_emergent_2015}
PF Machado, et~al., Emergent material properties of developing epithelial tissues.
\newblock {\em\protect\JournalTitle{BMC Biology}} \textbf{13}, 98 (2015).

\bibitem{brodland_differential_2002}
GW Brodland, The {Differential} {Interfacial} {Tension} {Hypothesis} ({DITH}): a comprehensive theory for the self-rearrangement of embryonic cells and tissues.
\newblock {\em\protect\JournalTitle{Journal of Biomechanical Engineering}} \textbf{124}, 188--197 (2002).

\bibitem{farhadifar_influence_2007}
R Farhadifar, JC Röper, B Aigouy, S Eaton, F Jülicher, The influence of cell mechanics, cell-cell interactions, and proliferation on epithelial packing.
\newblock {\em\protect\JournalTitle{Current biology: CB}} \textbf{17}, 2095--2104 (2007).

\bibitem{deeptissue}
D Haertter, Y Long, JM Crawford, CF Schmidt, DP Kiehart, Tracking and comprehending individual cell and junctional behavior in {{Drosophila}} dorsal closure using machine learning.
\newblock manuscript in preparation (2023).

\bibitem{PhysRevLett.123.058101}
X Li, A Das, D Bi, Mechanical heterogeneity in tissues promotes rigidity and controls cellular invasion.
\newblock {\em\protect\JournalTitle{Phys. Rev. Lett.}} \textbf{123}, 058101 (2019).

\bibitem{PERALTA20072583}
X Peralta, et~al., Upregulation of forces and morphogenic asymmetries in dorsal closure during drosophila development.
\newblock {\em\protect\JournalTitle{Biophysical Journal}} \textbf{92}, 2583--2596 (2007).

\bibitem{wells_complete_2014}
AR Wells, et~al., Complete canthi removal reveals that forces from the amnioserosa alone are sufficient to drive dorsal closure in {Drosophila}.
\newblock {\em\protect\JournalTitle{Molecular Biology of the Cell}} \textbf{25}, 3552--3568 (2014).

\bibitem{pmid30458138}
A Sumi, et~al., {{A}dherens {J}unction {L}ength during {T}issue {C}ontraction {I}s {C}ontrolled by the {M}echanosensitive {A}ctivity of {A}ctomyosin and {J}unctional {R}ecycling}.
\newblock {\em\protect\JournalTitle{Dev Cell}} \textbf{47}, 453--463 (2018).

\bibitem{pmid21516109}
R Levayer, A Pelissier-Monier, T Lecuit, {{S}patial regulation of {D}ia and {M}yosin-{I}{I} by {R}ho{G}{E}{F}2 controls initiation of {E}-cadherin endocytosis during epithelial morphogenesis}.
\newblock {\em\protect\JournalTitle{Nat Cell Biol}} \textbf{13}, 529--540 (2011).

\bibitem{Mateus}
AM Mateus, N Gorfinkiel, S Schamberg, A Martinez~Arias, Endocytic and recycling endosomes modulate cell shape changes and tissue behaviour during morphogenesis in drosophila.
\newblock {\em\protect\JournalTitle{PLoS One}} \textbf{6}, e18729 (2011).

\bibitem{yang_correlating_2017}
X Yang, et~al., Correlating cell shape and cellular stress in motile confluent tissues.
\newblock {\em\protect\JournalTitle{Proceedings of the National Academy of Sciences}} \textbf{114}, 12663--12668 (2017).

\bibitem{teomy_confluent_2018}
E Teomy, DA Kessler, H Levine, Confluent and nonconfluent phases in a model of cell tissue.
\newblock {\em\protect\JournalTitle{Physical Review E}} \textbf{98}, 042418 (2018) Publisher: American Physical Society.

\bibitem{arzash_tuning_2023}
S Arzash, I Tah, AJ Liu, ML Manning, Tuning for fluidity using fluctuations in biological tissue models.
\newblock (2023).

\bibitem{pmid10769037}
DP Kiehart, CG Galbraith, KA Edwards, WL Rickoll, RA Montague, {{M}ultiple forces contribute to cell sheet morphogenesis for dorsal closure in {D}rosophila}.
\newblock {\em\protect\JournalTitle{J Cell Biol}} \textbf{149}, 471--490 (2000).

\bibitem{pmid22404919}
A Sokolow, Y Toyama, DP Kiehart, GS Edwards, {{C}ell ingression and apical shape oscillations during dorsal closure in {D}rosophila}.
\newblock {\em\protect\JournalTitle{Biophys J}} \textbf{102}, 969--979 (2012).

\bibitem{SF9710500016}
PG de~Gennes, Possible experiments on two-dimensional nematics.
\newblock {\em\protect\JournalTitle{Symp. Faraday Soc.}} \textbf{5}, 16--25 (1971).

\bibitem{doi:10.1073/pnas.1705921114}
X Yang, et~al., Correlating cell shape and cellular stress in motile confluent tissues.
\newblock {\em\protect\JournalTitle{Proceedings of the National Academy of Sciences}} \textbf{114}, 12663--12668 (2017).

\bibitem{pmid26389664}
C Collinet, M Rauzi, PF Lenne, T Lecuit, {{L}ocal and tissue-scale forces drive oriented junction growth during tissue extension}.
\newblock {\em\protect\JournalTitle{Nat Cell Biol}} \textbf{17}, 1247--1258 (2015).

\bibitem{pmid21068726}
M Rauzi, PF Lenne, T Lecuit, {{P}lanar polarized actomyosin contractile flows control epithelial junction remodelling}.
\newblock {\em\protect\JournalTitle{Nature}} \textbf{468}, 1110--1114 (2010).

\bibitem{pmid18978783}
M Rauzi, P Verant, T Lecuit, PF Lenne, {{N}ature and anisotropy of cortical forces orienting {D}rosophila tissue morphogenesis}.
\newblock {\em\protect\JournalTitle{Nat Cell Biol}} \textbf{10}, 1401--1410 (2008).

\bibitem{pmid19879198}
R Fernandez-Gonzalez, SdeM Simoes, JC Röper, S Eaton, JA Zallen, {{M}yosin {I}{I} dynamics are regulated by tension in intercalating cells}.
\newblock {\em\protect\JournalTitle{Dev Cell}} \textbf{17}, 736--743 (2009).

\bibitem{lawson-keister_jamming_2021}
E Lawson-Keister, ML Manning, Jamming and arrest of cell motion in biological tissues.
\newblock {\em\protect\JournalTitle{Current Opinion in Cell Biology}} \textbf{72}, 146--155 (2021).

\bibitem{mongera_fluid--solid_2018}
A Mongera, et~al., A fluid-to-solid jamming transition underlies vertebrate body axis elongation.
\newblock {\em\protect\JournalTitle{Nature}} \textbf{561}, 401--405 (2018) Number: 7723 Publisher: Nature Publishing Group.

\bibitem{petridou_rigidity_2021}
NI Petridou, B Corominas-Murtra, CP Heisenberg, E Hannezo, Rigidity percolation uncovers a structural basis for embryonic tissue phase transitions.
\newblock {\em\protect\JournalTitle{Cell}} \textbf{184}, 1914--1928.e19 (2021).

\bibitem{Petridou2021-ow}
NI Petridou, B Corominas-Murtra, CP Heisenberg, E Hannezo, Rigidity percolation uncovers a structural basis for embryonic tissue phase transitions.
\newblock {\em\protect\JournalTitle{Cell}} \textbf{184}, 1914--1928.e19 (2021).

\bibitem{doi:10.1080/00018737100101261}
VK Shante, S Kirkpatrick, An introduction to percolation theory.
\newblock {\em\protect\JournalTitle{Advances in Physics}} \textbf{20}, 325--357 (1971).

\bibitem{PhysRevE.60.6361}
HP Hsu, MC Huang, Percolation thresholds, critical exponents, and scaling functions on planar random lattices and their duals.
\newblock {\em\protect\JournalTitle{Phys. Rev. E}} \textbf{60}, 6361--6370 (1999).

\bibitem{PhysRevE.80.041101}
AM Becker, RM Ziff, Percolation thresholds on two-dimensional voronoi networks and delaunay triangulations.
\newblock {\em\protect\JournalTitle{Phys. Rev. E}} \textbf{80}, 041101 (2009).

\bibitem{10.1073/pnas.1917853118}
J Devany, DM Sussman, T Yamamoto, ML Manning, ML Gardel, Cell cycle\textendash dependent active stress drives epithelia remodeling.
\newblock {\em\protect\JournalTitle{Proceedings of the National Academy of Sciences}} \textbf{118}, e1917853118 (2021).

\bibitem{Lan_2015}
H Lan, Q Wang, R Fernandez-Gonzalez, JJ Feng, A biomechanical model for cell polarization and intercalation during drosophila germband extension.
\newblock {\em\protect\JournalTitle{Physical Biology}} \textbf{12}, 056011 (2015).

\bibitem{lynch_cellular_2013}
HE Lynch, et~al., Cellular mechanics of germ band retraction in {Drosophila}.
\newblock {\em\protect\JournalTitle{Developmental Biology}} \textbf{384}, 205--213 (2013).

\bibitem{PhysRevLett.114.225501}
CP Goodrich, AJ Liu, SR Nagel, The principle of independent bond-level response: Tuning by pruning to exploit disorder for global behavior.
\newblock {\em\protect\JournalTitle{Phys. Rev. Lett.}} \textbf{114}, 225501 (2015).

\bibitem{hexner_role_2018}
D Hexner, AJ Liu, SR Nagel, Role of local response in manipulating the elastic properties of disordered solids by bond removal.
\newblock {\em\protect\JournalTitle{Soft Matter}} \textbf{14}, 312--318 (2018) Publisher: The Royal Society of Chemistry.

\bibitem{rocks_designing_2017}
JW Rocks, et~al., Designing allostery-inspired response in mechanical networks.
\newblock {\em\protect\JournalTitle{Proceedings of the National Academy of Sciences}} \textbf{114}, 2520--2525 (2017) Publisher: Proceedings of the National Academy of Sciences.

\bibitem{hagh_transient_2022}
VF Hagh, SR Nagel, AJ Liu, ML Manning, EI Corwin, Transient learning degrees of freedom for introducing function in materials.
\newblock {\em\protect\JournalTitle{Proceedings of the National Academy of Sciences}} \textbf{119}, e2117622119 (2022) Publisher: Proceedings of the National Academy of Sciences.

\bibitem{PhysRevX.11.021045}
M Stern, D Hexner, JW Rocks, AJ Liu, Supervised learning in physical networks: From machine learning to learning machines.
\newblock {\em\protect\JournalTitle{Phys. Rev. X}} \textbf{11}, 021045 (2021).

\bibitem{dillavouPRApplied2022}
S Dillavou, M Stern, AJ Liu, DJ Durian, Demonstration of {Decentralized} {Physics}-{Driven} {Learning}.
\newblock {\em\protect\JournalTitle{Physical Review Applied}} \textbf{18}, 014040 (2022) Publisher: American Physical Society.

\bibitem{z_drosophila_1991}
Z Kam, JS Minden, DA Agard, JW Sedat, M Leptin, Drosophila gastrulation: analysis of cell shape changes in living embryos by three-dimensional fluorescence microscopy.
\newblock {\em\protect\JournalTitle{Development (Cambridge, England)}} \textbf{112}, 365--370 (1991).

\bibitem{KIEHART200687}
DP Kiehart, et~al., {\em Chapter 9 - Ultraviolet Laser Microbeam for Dissection of Drosophila Embryos} ed.{} JE Celis.
\newblock (Academic Press, Burlington), Third edition edition, pp. 87--103 (2006).

\bibitem{goldstein_chapter_1994}
DP Kiehart, RA Montague, WL Rickoll, D Foard, GH Thomas, Chapter 26 {High}-{Resolution} {Microscopic} {Methods} for the {Analysis} of {Cellular} {Movements} in {Drosophila} {Embryos} XXmissing booktitle \& seriesXXMethods in {Cell} {Biology}, eds.{} LSB Goldstein, EA Fyrberg.
\newblock (Academic Press) Vol.{}~44, pp. 507--532 (1994) ISSN: 0091-679X.

\bibitem{oda_real-time_2001}
H Oda, S Tsukita, Real-time imaging of cell-cell adherens junctions reveals that {Drosophila} mesoderm invagination begins with two phases of apical constriction of cells.
\newblock {\em\protect\JournalTitle{Journal of Cell Science}} \textbf{114}, 493--501 (2001).

\bibitem{haertter_deepprojection_2022}
D Haertter, et~al., {DeepProjection}: specific and robust projection of curved {2D} tissue sheets from {3D} microscopy using deep learning.
\newblock {\em\protect\JournalTitle{Development}} \textbf{149} (2022) \_eprint: https://journals.biologists.com/dev/article-pdf/149/21/dev200621/2279764/dev200621.pdf.

\bibitem{10.1038/s41592-018-0261-2}
T Falk, et~al., U-{{Net}}: Deep learning for cell counting, detection, and morphometry.
\newblock {\em\protect\JournalTitle{Nature Methods}} \textbf{16}, 67--70 (2019).

\bibitem{celis_chapter_2006}
DP Kiehart, et~al., Chapter 9 - {Ultraviolet} {Laser} {Microbeam} for {Dissection} of {Drosophila} {Embryos} in {\em Cell {Biology} ({Third} {Edition})}, ed.{} JE Celis.
\newblock (Academic Press, Burlington), Third edition edition, pp. 87--103 (2006).

\bibitem{10.2976/1.2955565}
A Rodriguez-Diaz, et~al., Actomyosin purse strings: {{Renewable}} resources that make morphogenesis robust and resilient.
\newblock {\em\protect\JournalTitle{HFSP Journal}} \textbf{2}, 220--237 (2008).

\bibitem{10.1091/mbc.e14-07-1190}
AR Wells, et~al., Complete canthi removal reveals that forces from the amnioserosa alone are sufficient to drive dorsal closure in {{Drosophila}}.
\newblock {\em\protect\JournalTitle{Molecular Biology of the Cell}} \textbf{25}, 3552--3568 (2014).

\bibitem{SUSSMAN2017400}
DM Sussman, cellgpu: Massively parallel simulations of dynamic vertex models.
\newblock {\em\protect\JournalTitle{Computer Physics Communications}} \textbf{219}, 400--406 (2017).

\end{thebibliography}


\begin{thebibliography}{10}

\bibitem{doi:10.1080/00107518408210979}
D Weaire, N Rivier, Soap, cells and statistics—random patterns in two
  dimensions.
\newblock {\em\protect\JournalTitle{Contemporary Physics}} \textbf{25}, 59--99
  (1984).

\bibitem{PhysRevE.63.011402}
F Graner, Y Jiang, E Janiaud, C Flament, Equilibrium states and ground state of
  two-dimensional fluid foams.
\newblock {\em\protect\JournalTitle{Phys. Rev. E}} \textbf{63}, 011402 (2000).

\bibitem{pmid26237129}
JA Park, et~al., {{U}njamming and cell shape in the asthmatic airway
  epithelium}.
\newblock {\em\protect\JournalTitle{Nat Mater}} \textbf{14}, 1040--1048 (2015).

\bibitem{Atia_et_al}
L Atia, et~al., Geometric constraints during epithelial jamming.
\newblock {\em\protect\JournalTitle{Nature Physics}} \textbf{14}, 613--620
  (2018).

\bibitem{Wang13541}
X Wang, et~al., Anisotropy links cell shapes to tissue flow during convergent
  extension.
\newblock {\em\protect\JournalTitle{Proceedings of the National Academy of
  Sciences}} \textbf{117}, 13541--13551 (2020).

\bibitem{10.1073/pnas.1917853118}
J Devany, DM Sussman, T Yamamoto, ML Manning, ML Gardel, Cell cycle\textendash
  dependent active stress drives epithelia remodeling.
\newblock {\em\protect\JournalTitle{Proceedings of the National Academy of
  Sciences}} \textbf{118}, e1917853118 (2021).

\bibitem{doi:10.1091/mbc.E21-11-0537}
RP Moore, et~al., Superresolution microscopy reveals actomyosin dynamics in
  medioapical arrays.
\newblock {\em\protect\JournalTitle{Molecular Biology of the Cell}}
  \textbf{33}, ar94 (2022) PMID: 35544300.

\bibitem{farhadifar_influence_2007}
R Farhadifar, JC Röper, B Aigouy, S Eaton, F Jülicher, The influence of cell
  mechanics, cell-cell interactions, and proliferation on epithelial packing.
\newblock {\em\protect\JournalTitle{Current biology: CB}} \textbf{17},
  2095--2104 (2007).

\bibitem{brodland_differential_2002}
GW Brodland, The {Differential} {Interfacial} {Tension} {Hypothesis} ({DITH}):
  a comprehensive theory for the self-rearrangement of embryonic cells and
  tissues.
\newblock {\em\protect\JournalTitle{Journal of Biomechanical Engineering}}
  \textbf{124}, 188--197 (2002).

\bibitem{Bi_nature_physics}
D Bi, JH Lopez, JM Schwarz, ML Manning, A density-independent rigidity
  transition in biological tissues.
\newblock {\em\protect\JournalTitle{Nature Physics}} \textbf{11}, 1074--1079
  (2015).

\bibitem{PhysRevLett.97.170201}
E Bitzek, P Koskinen, F G\"ahler, M Moseler, P Gumbsch, Structural relaxation
  made simple.
\newblock {\em\protect\JournalTitle{Phys. Rev. Lett.}} \textbf{97}, 170201
  (2006).

\bibitem{de_gennes_physics_1993}
P de~Gennes, J Prost, {\em The {Physics} of {Liquid} {Crystals}}, International
  series of monographs on physics.
\newblock (Clarendon Press), (1993).

\bibitem{C3SM52323C}
G Duclos, S Garcia, HG Yevick, P Silberzan, Perfect nematic order in confined
  monolayers of spindle-shaped cells.
\newblock {\em\protect\JournalTitle{Soft Matter}} \textbf{10}, 2346--2353
  (2014).

\bibitem{doi:10.1073/pnas.1707210114}
X Li, et~al., On the mechanism of long-range orientational order of
  fibroblasts.
\newblock {\em\protect\JournalTitle{Proceedings of the National Academy of
  Sciences}} \textbf{114}, 8974--8979 (2017).

\bibitem{SF9710500016}
PG de~Gennes, Possible experiments on two-dimensional nematics.
\newblock {\em\protect\JournalTitle{Symp. Faraday Soc.}} \textbf{5}, 16--25
  (1971).

\bibitem{PhysRevLett.123.058101}
X Li, A Das, D Bi, Mechanical heterogeneity in tissues promotes rigidity and
  controls cellular invasion.
\newblock {\em\protect\JournalTitle{Phys. Rev. Lett.}} \textbf{123}, 058101
  (2019).

\end{thebibliography}
\end{document}


\section*{Supplementary Material:\\ Minimal vertex model explains how the amnioserosa avoids fluidization during \textit{Drosophila} dorsal closure}
\subsection{Short introduction to vertex models}
\label{SI_intro_vertex_models}
Vertex models for tissues were motivated by two-dimensional models of dry foams made up of polygonal bubbles. In such foams, the vertices of bubbles tend to adjust their positions to balance forces arising from line tensions at the boundaries between adjacent bubbles; this force balance is achieved by minimizing the energy cost associated with bubble-bubble boundaries \cite{doi:10.1080/00107518408210979,PhysRevE.63.011402}. 
Tissue imaging typically provides time-evolving 2D projections of the tissue layers. 2D modeling of 2D projections in vertex models is appropriate, since tension is provided  by a 1D ring of acto-myosin, and adhesion is provided by E-cadherin also in a ring-like geometry. One can visualize such a model as a tightly packed layer of elastic polygons, having a finite relaxed area, with a line tension around the circumference and "slippery" adhesiveness to their neighbors that prevents gaps, but allows sliding. With these assumptions, computer simulations show that barriers preventing cells from sliding past each other are low when cells are elongated, i.e., the tissue reacts like a fluid to external forces on time scales determined by the remaining friction. Barriers are high, in contrast, when cells are more rounded, i.e., the tissue reacts like a solid under external forces in the sense that the tissue may deform, but cells maintain their neighbors.  This type of very simple modeling has been extraordinarily successful in providing quantitative predictions of the decrease and loss of solid character in a wide range of epithelial tissues \cite{pmid26237129,Atia_et_al,Wang13541,10.1073/pnas.1917853118} and monolayer cell cultures. It was found that one parameter, mean cell shape index  ($\bar q = \langle p_i/\sqrt a_i \rangle$ ), where $p$ is the cell perimeter and $a$ is its area, was dominant in determining phase state in many systems \cite{pmid26237129,Atia_et_al,Wang13541}. 
Within the standard vertex model, the AS is a two-dimensional plane comprising irregular polygons that represent cells and that display no overlaps or gaps, as shown in Fig.\,~1A,B  in the main text. The positions of the vertices adjust to minimize the mechanical energy
\begin{equation}
E = \sum^{N}_{i=1} \frac{1}{2}k_a (a_i - a_{0})^2 + \frac{1}{2}k_p (p_i - p_{0,i})^2.
\label{eq:energy}
\end{equation}
Here $N$ is the total number of cells, $p_i$ and $a_i$ are the actual perimeter and area of cell $i$ and $p_{0,i}$ and $a_{0}$ are its preferred perimeter and area. $k_a$ and $k_p$ represent the area and perimeter moduli of the cells. The first term in Eq.~\ref{eq:energy} represents the area elasticity of the cells. According to this term, there are restoring forces if the area of cell $i$, $a_i$, differs from its preferred area, $a_{0}$. We assume that the cell volumes are constant due to incompressibility, changes of area can arise from changes in the height of cells. We note that there are medioapical arrays of actomyosin that condense and relax as pulsing AS cells contract and expand~\cite{doi:10.1091/mbc.E21-11-0537}. In the vertex model, the medioapical arrays would control the preferred area, $a_{0,i}$. In our model, we neglect such effects, taking $a_{0,i}$ to be constant in time and the same for every cell ($a_{0,i}=a_0=\mathrm{const}$).  

The second term in Eq.~\ref{eq:energy}, involving the cell perimeter, originates from the sum of two contributions 
\begin{equation}
    k_{p,i} p_i^2 + \gamma_i p_i = \frac{1}{2}k_p(p_i - p_{0,i})^2 
\label{eq:perimeter_energy}
\end{equation}
where $p_{0,i} = -\gamma_i (2k_{p,i})$ is the effective target cell perimeter \cite{Bi_nature_physics}. The first term on the left side of Equ. \ref{eq:perimeter_energy} approximates active contractility of the actomyosin sub-cellular cortex. According to this term, the cortex acts like a spring that prefers each cell to have a perimeter of zero (complete contraction); this contributes a term proportional to the square of the perimeter~\cite{farhadifar_influence_2007}. The second contribution (second term on left side of Eq. \ref{eq:perimeter_energy}) arises from cell-cell adhesion and cortical tension which are proportional to the perimeter. They combine to give an effective line tension $\gamma_i$ that penalizes a nonzero perimeter if $\gamma_i>0$ (if contractility dominates) or rewards a nonzero perimeter if $\gamma_i<0$ (if cell-cell adhesion dominates)~\cite{farhadifar_influence_2007,brodland_differential_2002,Bi_nature_physics}. 

\subsection{Details of the amnioserosa vertex model}
\label{SI_model_details}

The initial AS cell sheet is modeled as a two-dimensional rectangular plane sheet with box size  $L_x \times L_{y,\mathrm{initial}}$ that is tiled with $N_\mathrm{initial}=256$ cells with no gaps or overlaps. We use periodic boundary conditions throughout the process. For simplicity, we set the preferred area for each cell $i$ to be the same for all cells: $a_{0,i}=1$ .

The experimentally observed heterogeneity in cell shape index of the AS tissue is introduced in our model through the preferred perimeters of cells ($p_{0,i}$ for cell $i$). The initial values of $p_{0,i}$ are drawn from a Gaussian distribution with mean $\mu = 4.24$ and standard deviation $\sigma = 0.45$. We then minimize the energy using the FIRE algorithm \cite{PhysRevLett.97.170201} until the residual force on each vertex drops below $10^{-6}$ to obtain the initial state of the tissue. After minimization, we compare the distribution of shape index from both simulation and experiment (Fig.\,1C). 

We approximate the dorsal closure process by applying uni-axial quasi-static compression to our model tissue. We hold $L_x$ fixed and with each step, we decrease $L_y$ by $\epsilon = 0.01$ and re-equilibrate the system (minimize the mechanical energy) so that there is force balance at every vertex. 

During the process, we also decrease all of the preferred cell perimeters with a linear rate corresponding to the average perimeter drop observed in the experiments ($m_p=2.76$) (Fig.\,1E):
\begin{equation}
\label{eq:Peri_dec}
  p_{0,i}(\Delta A(t)) =  p_{0,i}(\Delta A(0)) - m_p \Delta A(t).
\end{equation}

\subsection{Relation between AS height and AS area}
\label{SI_AS_height_area}
\begin{figure}[ht]
\center
\includegraphics[width=0.4\textwidth]{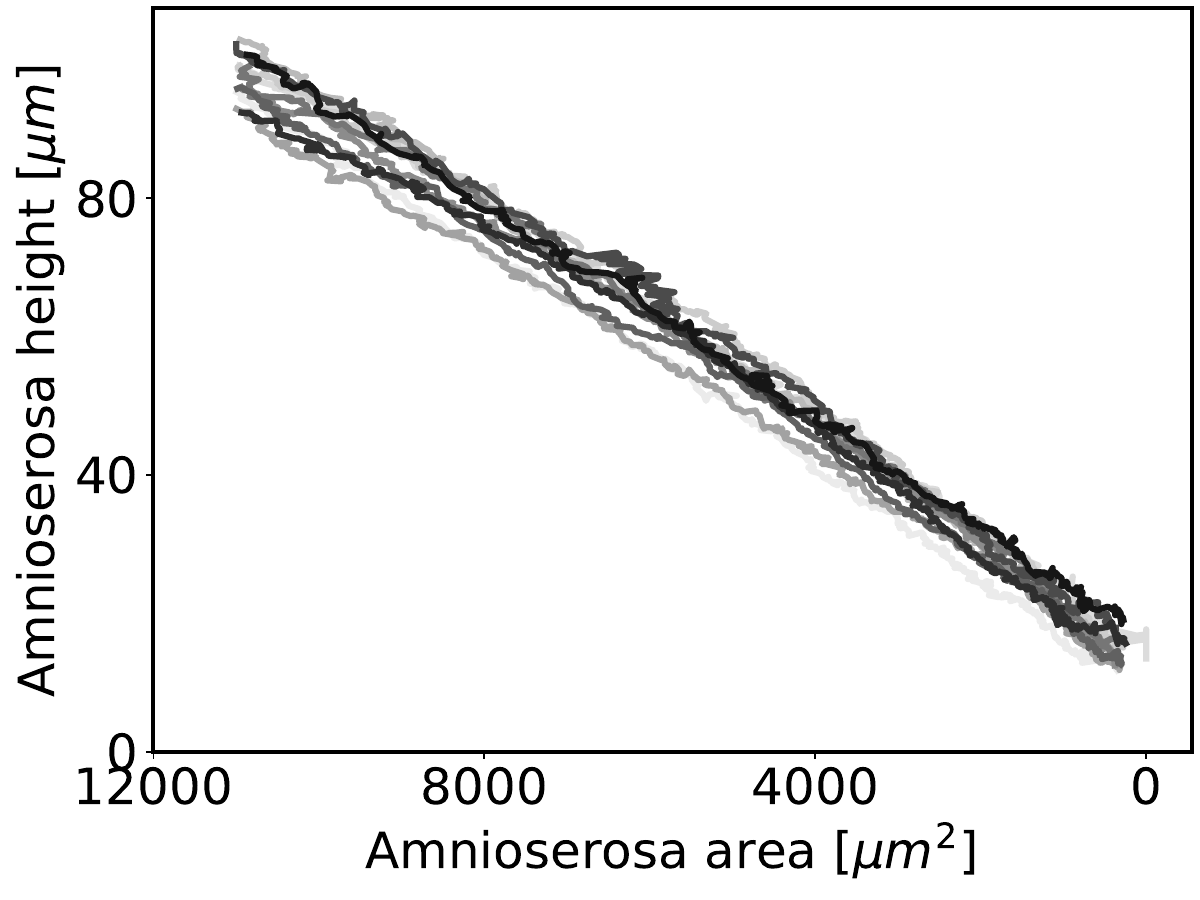}
\caption{\textbf{Linear relation between the AS height and area.} Each color represents one embryo (N=12).}
  \label{SI_Fig_ARea_vs_height}
\end{figure}

In Fig.\,\ref{SI_Fig_ARea_vs_height} relation between the AS height and area is demonstrated. 
\subsection{Cell shape index and aspect ratio} 
\label{SI_shape_index}
The shape index of cell $i$ is defined as $q_i \equiv  p_i/\sqrt{a_i}$, where $p_i$ is the cell perimeter and $a_i$ is the cell area. The mean cell shape index is $(\bar{q})=\langle \frac{p_i}{\sqrt{a_i}} \rangle=\frac{1}{N}\sum_{i=1}^N \frac{p_i}{\sqrt{a_i}}$.

The aspect ratio of a cell is defined as 
\begin{equation}
{\alpha} = \frac {\sqrt {\frac{1}{\lambda_1}}}{\sqrt {\frac{1}{\lambda_2}}}
\label{eq:AR}
\end{equation}
where $\lambda_1$ and $\lambda_2$ are smallest and largest eigenvalues of the shape tensor $I$, defined as 
\begin{equation}
I = 
\begin{pmatrix}
I_{XX} & I_{XY} \\
I_{XY} & I_{YY} 
\end{pmatrix}
\label{eq:moment}
\end{equation}
Here, $I_{XX} = \iint_R y^2 dA$, $I_{YY} = \iint_R x^2 dA$, and $I_{XY} = \iint_R xy dA$. The tensor $I$, calculated as in Ref.~\cite{Atia_et_al}, corresponds to the moment-of-inertia tensor in the case where the mass density per unit area is constant. Thus $I$ measures the second area moment of the polygon representing the apical surface of the cell, relative to a fixed point, weighting each infinitesimal area element of the cell equally. In this case, the fixed point is the center of the cell, as defined by the first moment (the mean) of the area distribution of the polygon. Thus, the second moment measures the variance of the area distribution. In 2-dimensions, the tensor is a $2 \times 2$ matrix as defined above and its two eigenvalues $\lambda_1$ and $\lambda_2$ along two directions that in the case of an ellipse would correspond to the minor and major axes, respectively.

\subsection{Cell orientational order parameter}
\label{SI_orientational_order_parameter}
During dorsal closure, cells tend to align along the anterior-posterior axis. According to the standard vertex model, this alignment should enhance the solid character of the tissue~\cite{Wang13541}. To better understand the relationship of solidity to the collective alignment among aminoserosa cells, we quantify their degree of orientational order~\cite{de_gennes_physics_1993,C3SM52323C,doi:10.1073/pnas.1707210114}. This tells us the degree to which cells align relative to the anterior-posterior axis. The 2D orientational order parameter $\bar Q$ is \cite{SF9710500016} 
\begin{equation}
\bar Q = \frac{1}{N} \sum_{i=1}^{N} \cos (2\theta_i)
\label{eq:oop}
\end{equation}
where $\theta$ is the angle of the major axis of the cell relative to the anterior-posterior axis. As described in Sec\,\ref{SI_shape_index}, the direction of the cell's major axis is given by the largest eigenvector of the shape tensor $I$. Note $\bar Q=0$ when cells are randomly aligned in the tissue (the isotropic case) and $\bar Q = 1$ if all cells are perfectly aligned along the anterior-posterior axis.

\subsection{Relation between mean and standard deviation of cell aspect ratio}
\label{SI_AR_STD_mean}
\begin{figure}[H]
\center
\includegraphics[width=0.6\textwidth]{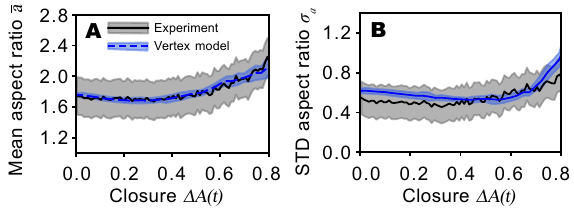}
\caption{\textbf{Comparison of mean (A) and standard deviation (B) of cell aspect ratio between experimental data (black) and model (blue) during dorsal closure.} Lines show mean, and shaded areas standard deviation of various initial configurations (model, N = 10) or different embryos (experiment, N = 12).}
\label{SI_Fig_AR}
\end{figure}

Similar to the cell shape index $q$, the cellular aspect ratio $\alpha$ also characterizes cell shape \cite{Atia_et_al}. Fig.\,\ref{SI_Fig_AR} demonstrates that our model successfully captures experimental observations for both the mean and standard deviations of the tissue aspect ratio.

\subsection{Rigidity percolation in uniaxially deformed tissue}
\label{SI_percolation}

Uniaxial deformation during dorsal closure produces anisotropy in the tissue which can affect the percolation threshold. To probe how the percolation threshold $f_c$ depends on the uniaxial deformation, we take the model tissue at each value of $\Delta A(t)$ and randomly assign each cell-cell junction to have nonzero tension with probability $f$ (Fig.\,\ref{fig:SI_percolation}A). Percolation of the system as a whole requires percolation in both the x and y direction. Therefore, the percolation threshold of the system is given by $f_c=\mathrm{max}[f_c^{x},f_c^y]$ (Fig.\,\ref{fig:SI_percolation}C,D). To determine the percolation thresholds $f_c^{x}$ and $f_c^y$ at different stages of closure $\Delta A$, we each assessed the probability of obtaining a system-spanning connected path of non-zero tension edges $P_{tension}^{X,Y}$ for different ratios of junctions with non-zero tension $f_r$ (Fig.\,\ref{fig:SI_percolation}C,D). The respective percolation thresholds $(f_c) : (\frac{d^2 P_{tension}}{df_r^2} = 0 )$ were the inflection points of $P_{tension}$. In Fig.\, \ref{fig:SI_percolation}B, note that $f_c^x \approx 0.66$ remains unchanged but $f_c^y$ decreases with increasing $\Delta A(t)$. As a result, $f_c=\mathrm{max}[f_c^{x},f_c^y]$ remains constant at $f_c \approx 0.66$ throughout dorsal closure. Thus we find that the fraction of junctions with non-zero tension must satisfy $f_r \ge f_c \approx 0.66$ for solid behavior. 

Uniaxial deformation can also affect $f_r$, the fraction of junctions with non-zero tension. To determine the fraction of junctions with non-zero tension in the model, we chose a threshold corresponding to the noise floor of the tension \cite{PhysRevLett.123.058101}, so that junctions with $\tau_J> 10^{-4}$ are counted as having finite tension. In the experiments, we found that the initial recoil velocity $v_r$ was independent of junction straightness $S$ for $S \lesssim 0.93$ (Fig.\,2E), but rose linearly with $S$ above this threshold. We therefore assume that only junctions with $S>0.93$ carry tension and contribute to solid response. In Fig.\,~\ref{fig:SI_percolation}B it is evident that $f_r>f_c$ throughout the process, consistent with the system being solid.

In summary, the AS tissue maintains a percolating network of tense cell-cell junctions across the dorsal opening during the entire process of closure, consistent with its solid character.

The decrease of preferred cell perimeter is crucial. If we leave the preferred perimeter fixed at its initial value in our model, we obtain $f_r$ as shown in Fig.\,\ref{fig:SI_percolation}B. Clearly, $f_r$ falls below $f_c$  so that the system fluidizes.

\subsection{Mean junction tension in tissues with and without orientational alignment of the cells}
\label{SI_anisotropy_alignment}
\begin{figure}[h]
\centering
\includegraphics[width=0.35\textwidth]{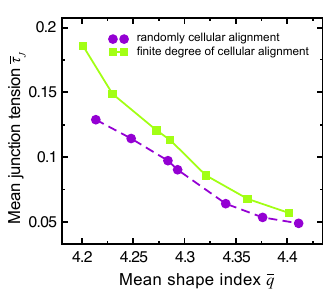}
\caption{\textbf{Comparison of mean junction tension $\bar \tau_J$ for tissues with and without cellular orientational alignment at different mean cell shape indices $\bar q$.}}
\label{SI_Fig_cortical_tension_alignment}
\end{figure}

To understand the effect of the orientational alignment of cells on mechanical stiffness in our model, we compared our model, where we found orientational alignment of cells  (Fig.\,2C) during uni-axial constriction, with simulations with randomly aligned cells. We then quantified the mean junction tension $\Bar \tau_J$, as a measure for tissue stiffness, in model configurations with different mean shape indices $\bar q$. For a given mean shape index, we find that the junction tension is higher if the cells are oriented than if they are not (Fig.\,\ref{SI_Fig_cortical_tension_alignment}), consistent with Ref.~\cite{Wang13541}. 

\subsection{Effect of uniaxial deformation on tissue phase state}
\label{SI_isotropic_constriction}

\begin{figure}[H]
\center
\includegraphics[width=0.6\textwidth]{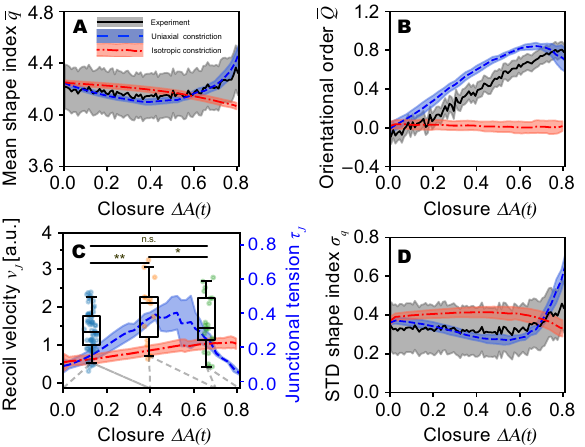}
\caption{\textbf{Comparison of experimental data (black) with model results with uniaxial deformation (blue) and isotropic deformation (red) during dorsal closure.} \textbf{(A)} Time course of mean shape index $\bar q$. \textbf{(B)} Time course of average orientational order parameter $\bar{Q}$. \textbf{(C)} Box plot of initial junction recoil velocity of junction vertices after performing laser ablation of the junction, and time-course of average junction tension $\bar{\tau_c}$. Boxplots show data for three intervals of $\Delta A$ ($\Delta A<0.4$, $0.4 \leq \Delta A < 0.7$, $\Delta A \geq 0.7$). Whiskers extend to the 5th and 95th percentiles, while the boxes delineate the interquartile range, and the horizontal lines within the boxes indicate the median values. An ANOVA followed by a post-hoc Tukey's HSD test was conducted to assess statistical significance (*: $p<0.1$, **: $p<0.05$). We performed and evaluated cuts of $N=97$ junctions.  \textbf{(D)} Time-course of standard deviation of shape index ($\sigma_q$). Lines show mean, and equally color intervals show respective standard deviations (N = 10 model initial configurations, N = 12 embryos).}
\label{SI_Fig_isotropic}
\end{figure}

We studied the effect of uniaxial deformation of the AS, as implemented in our model, in contrast to isotropic deformation. We progressively decreased the tissue size in our model in both x and y direction, and compared it with our model results for uniaxial deformation at each given $\Delta A$ in terms of  mean shape index $\bar{q}$ (Fig.\,\ref{SI_Fig_isotropic}A), average orientational order $\bar Q$  (Fig.\,\ref{SI_Fig_isotropic}B), average junction tension $\bar{\tau_J}$ and initial recoil velocity $v_J$ (Fig.\,\ref{SI_Fig_isotropic}C), and standard deviation of shape index $\sigma_q$ (Fig.\,\ref{SI_Fig_isotropic}D). We found that uniaxial deformation of the AS in our is crucial to recapitulate the experimentally measured time courses of cell-shape features.

\begin{figure}[H]
\centering
\includegraphics[width=0.66\textwidth]{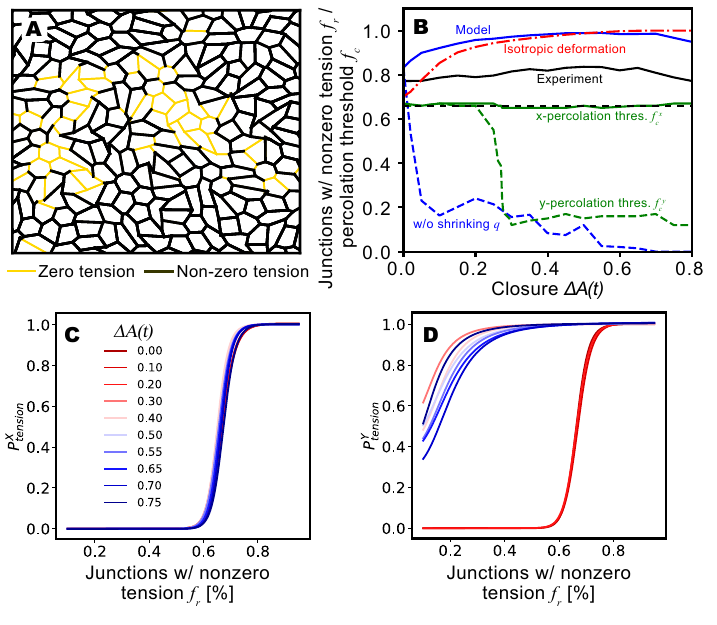}
\caption{\textbf{Analysis of percolation in junction network.} (\textbf{A}) Snapshot of tension network percolation: Black thick lines are edges with finite tensions (percolating throughout the tissue), while other edges have $\tau = 0$. (\textbf{B}) Fraction of junctions with nonzero tension, $f_r$, in experiment (black solid) and model (blue solid) as a function of $\Delta A(t)$. The model predicts the minimum fraction $f_c$ needed for percolation of junctions with nonzero tension in the $x$ (green solid) and $y$ (green dashed) directions. In both experiment and model, $f_r>f_c={\mathrm{max}[f_c^x, f_c^y}]$, indicating that the system is solid. Black dashed line shows the effective bond percolation threshold $f_c^* = 0.66 = {\mathrm{max}[f_c^x, f_c^y}]$ above which the system is solid. If cell perimeters, and accordingly the shape indices of cells are held constant instead of shrinking with increasing $\Delta A(t)$, our model predicts that $f_r$ drops below the percolation threshold (blue dashed), leading to fluidization. If we apply isotropic instead of uni-axial deformation, $f_r$ (red dashed) remains above the percolation threshold. (\textbf{C}) Probability of obtaining percolation of cell edges with nonzero tensions ($P_{tension}^X$) vs. $f_r$ in the X directions (left) of our model tissue, at various closure stages (legend), which correspond to various tissue anisotropy stages. The point of inflection is the respective percolation threshold $f_c$. (\textbf{D}) Same plot as \textbf{C} for Y direction, showing substantially lower percolation thresholds than for X direction at late closure stages.}
\label{fig:SI_percolation}
\end{figure}

To test the effect of uniaxial vs. isotropic deformation on the solid phase of the AS, we assessed the fraction of junctions with nonzero tension $f_r$ during closure (red dashed line in Fig.\,\ref{fig:SI_percolation}B). For isotropic deformation, our model predicts $f_r$ well above the percolation threshold $f_c$, so non-zero tension junctions percolate in either case (isotropic and uni-axial).

\bibliography{Dorsal_Closure_2} 
\bibliographystyle{ieeetr}


\newpage

\newpage

\newpage

\newpage

\newpage

\newpage
\begin{figure}[H]
\centering
\includegraphics[width=0.7\textwidth]{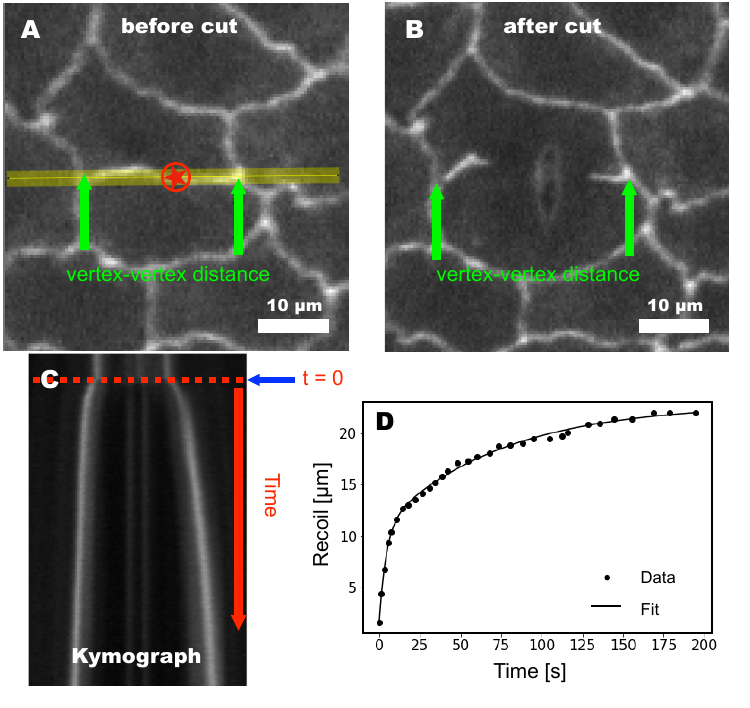}
\caption{\textbf{Measurement and evaluation of initial recoil velocity of vertices after laser cut.} (\textbf{A}),(\textbf{B}) Confocal fluorescent images of AS cells. The red star in (A) marks the ablation region. Green arrows show the vertex positions just prior to (A) and after (B) ablation. The yellow line (A) along which the motion of vertices is assessed using kymograph analysis. (\textbf{C}) Kymograph of positions of the two vertices  showing recoil after ablation (the red dotted line shows time of cut at $t=0$). (\textbf{D}) Time-course of recoil after ablation. The recoil was well fitted by a double exponential function $a_0 \exp(b_0t) + c_0 \exp(d_0t) +e_0$.}
\label{SI_Fig_cut}
\end{figure} 
\newpage